\documentclass[]{acmart}
\usepackage{multirow}
\AtBeginDocument{%
  \providecommand\BibTeX{{%
    \normalfont B\kern-0.5em{\scshape i\kern-0.25em b}\kern-0.8em\TeX}}}

\copyrightyear{2021}
\acmYear{2021}
\setcopyright{acmcopyright}\acmConference[Asian CHI Symposium 2021 ]{Asian CHI Symposium 2021 }{May 8--13, 2021}{Yokohama, Japan}
\acmBooktitle{Asian CHI Symposium 2021 (Asian CHI Symposium 2021 ), May 8--13, 2021, Yokohama, Japan}
\acmPrice{15.00}
\acmDOI{10.1145/3429360.3468180}
\acmISBN{978-1-4503-8203-8/21/05}

\begin{document}

\title{Changing Computer-Usage Behaviors: What Users Want, Use, and Experience}

\author{Mina Khan}
\authornote{\textbf{All authors contributed equally to this research. Inquiries may be directed to either: minakhan@media.mit.edu, eglassman@g.harvard.edu, or zeelpatel@college.harvard.edu}}
\affiliation{%
  \institution{M.I.T Media Lab, Massachusetts Institute of Technology}
  \city{Cambridge, MA}
  \country{Cambridge, MA, USA}}
\email{minakhan01@gmail.com}

\author{Zeel Patel}
\authornotemark[1]
\affiliation{%
  \institution{Department of Computer Science, Harvard University}
  \city{Cambridge}
  \state{MA}
  \country{Cambridge, MA, USA}
}
\email{zeelpatel@college.harvard.edu}

\author{Kathryn Wantlin}
\authornotemark[1]
\affiliation{%
  \institution{Department of Computer Science, Harvard University}
  \city{Cambridge}
  \state{MA}
  \country{Cambridge, MA, USA}
}
\email{kathrynwantlin@college.harvard.edu}

\author{Elena Glassman}
\authornotemark[1]
\affiliation{%
  \institution{Department of Computer Science, Harvard University}
  \city{Cambridge}
  \state{MA}
  \country{Cambridge, MA, USA}
}
\email{glassman@seas.harvard.edu}

\author{Pattie Maes}
\authornotemark[1]
\affiliation{%
  \institution{M.I.T Media Lab, Massachusetts Institute of Technology}
  \city{Cambridge, MA}
  \country{Cambridge, MA, USA}}
\email{pattie@media.mit.edu}

\renewcommand{\shortauthors}{Khan, et al.}

\begin{abstract}
Technology based screentime — the time an individual spends engaging with their computer or cell phone — has increased exponentially over the past decade, but perhaps most alarmingly amidst the COVID-19 pandemic. Although many software based interventions exist to reduce screentime, users report a variety of issues relating to the timing of the intervention, the strictness of the tool, and it’s ability to encourage organic, long-term habit formation. We develop guidelines for the design of behaviour intervention software by conducting a survey to investigate three research questions and further inform the mechanisms of computer-related behavior change applications. RQ1: What do people want to change and why/how? RQ2: What applications do people use or have used, why do they work or not, and what additional support is desired? RQ3: What are helpful/unhelpful computer breaks and why? Our survey had 68 participants and three key findings. First, time management is a primary concern, but emotional and physical side-effects are also important. Second, site blockers, self-trackers, and timers are commonly used, but they are ineffective as they are easy-to-ignore and not personalized. Third, away-from-computer breaks, especially involving physical activity, are helpful, whereas on-screen breaks are unhelpful, especially when they are long, because they are not refreshing. We recommend personalized and closed-loop computer-usage behavior change support and especially encouraging off-the-computer computer breaks.
\end{abstract}



\maketitle

\section{Introduction}

Technology has become an intrinsic part of our lives, but excessive technology use is also connected to several physical and psychological problems \cite{healy1999failure, subramani2017smartphone, soliman2020smartphone, kuss2018internet}.
Knowledge workers heavily rely on computers \cite{kim_understanding_2019} and recently, there has been a significant increase in computer usage, especially during the COVID-19 pandemic \cite{clement2020coronavirus}. 

There are several scales for accessing computer usage, e.g., computer use scale \cite{panero1997part}, attitudes toward computer usage scale \cite{popovich1987development}, and compulsive internet use scale \cite{meerkerk2009compulsive}. 
Research has been done to understand as well as modify computer and phone usage, e.g., encourage physical activity \cite{cambo2017breaksense}, and enable self-tracking \cite{rooksby_personal_2016} and better focus \cite{borghouts_timetofocus_2020, mark2017blocking}. There has also been research on identifying user needs, e.g., self-monitoring \cite{meyer_design_2017} and productivity needs \cite{guillou_is_2020, kim_understanding_2019}, and what people consider to be work-breaks \cite{epstein2016taking} and what breaks are helpful for productivity \cite{epstein2016taking}. 

We investigate 3 research questions to further inform the design of computer-usage behavior change applications, including applications for encouraging computer-related breaks. \textbf{RQ1. Computer-related Behavior Change Needs and Desired Changes:} What what do people want to change about their computer usage and why/how? \textbf{RQ2. Currently-used Computer-usage Behavior Change Tools and Further Needs:} What techniques or applications do people already use or have used for computer-related behavior change and why do they work or not work, and what kind of additional support do people require? \textbf{RQ3. Computer-related Helpful and Unhelpful Breaks:} What are helpful/unhelpful computer breaks and why are they are helpful/unhelpful?

\section{Related Work}

Research has focused on both understanding and modifying computer and phone usage as well as inferring user states and contexts based on their phone and computer usage, especially for delivering interventions. 
While previous work focused on specific user needs, e.g., promoting productivity, physical activity, self-tracking, and focus, 
our work sheds light on the overall computer-usage behavior change needs of the users and highlights what applications and experiences are helpful/unhelpful for the users and why.
Our related work is as follows.

\subsection{Understanding computer usage and user needs}
There are different scales for accessing computer usage, e.g., computer use scale \cite{panero1997part}, attitudes toward computer usage scale  \cite{popovich1987development}, and compulsive internet use scale \cite{meerkerk2009compulsive}. 
Researchers have also used in situ studies to investigate computer-usage behavior change needs, e.g., self-monitoring \cite{meyer_design_2017} and break prompts that discourage sedentary behavior \cite{luo_time_2018}. 
Researchers have also used diary studies and experience sampling to study more holistic scenarios, e.g., combining classic productivity with well-being \cite{guillou_is_2020}, understanding personal productivity beyond work-related productivity \cite{kim_understanding_2019}, and understanding what makes smartphone use meaningful or meaningless \cite{lukoff_what_2018}.
Epstein et al., in particular, conducted a survey to identify what types of breaks, e.g., digital and biological breaks, people consider as breaks from work and what are the desirable qualities of a  break, e.g., refreshing, relaxing \cite{epstein2016taking}. Epstein et al.'s diary study focused on helpful breaks for productivity but not necessarily helpful and unhelpful breaks for overall user needs, e.g., physical and emotional health \cite{epstein2016taking}.
Our work aims extend this work by surveying the overall user needs, identifying what support works and does not work for users, and which computer breaks are helpful/unhelpful for overall user needs.

\subsection{Computer-usage self-tracking and behavior change applications}
There have been several applications to help users monitor and manage their phone and computer usage. 
While some applications enable passive self-tracking \cite{barata_appinsight_2012, hu_screentrack_2019, rooksby_personal_2016}, others employ active interventions, e.g., for self-control on Facebook \cite{lyngs2020just}, for promoting mobility during work-breaks \cite{cambo2017breaksense}, for regulating phone usage \cite{kim_goalkeeper_2019, kim_lockntype_2019, okeke_good_2018}, and for blocking distractions to improve workplace focus and productivity \cite{mark2017blocking}.
Researchers have also studied different design choices, e.g., comparing goal-prompt versus removing newsfeed on Facebook \cite{lyngs2020just}, using physiological and location sensing for mobility prompts \cite{cambo2017breaksense}, comparing a point-of-choice prompt with an always-on progress bar to change sedentary behavior \cite{wang_point--choice_2019}, giving feedback on interruption durations to discourages distractions and interruptions \cite{borghouts_timetofocus_2020}, and using lockout mechanisms \cite{kim_goalkeeper_2019, kim_lockntype_2019} or even nudge-like vibrations \cite{okeke_good_2018} for regulating phone usage.
Researchers have also investigated individual differences in the effects of blocking workplace distractions \cite{mark2018effects}.
However, researchers focus on specific needs, e.g., increasing productivity and reducing distractions or sedentary behaviors, not on overall user needs.

\subsection{Modeling user states and opportune moments}
Studies have monitored phone and computer usage, even combined with physiological data, to not only automatically recognize breaks and work activities \cite{di_lascio_multi-sensor_2020}, but also to model opportune moments for transitions and breaks at work for optimizing happiness and productivity \cite{kaur_optimizing_2020}.
There is also research to infer opportune moments for well-being messages on mobile phones, e.g., interventions for attention management \cite{cavdar_multi-perspective_2020} and for discouraging sedentary behavior \cite{choi_multi-stage_2019}.
We focus on surveying the helpful/unhelpful breaks and support needs of users to further define the design of computer-usage behavior change interventions.

\section{Computer-Usage Behavior Change Needs Survey Design}

We conducted an anonymous survey and recruited the participants using convenience and snowball sampling. We shared the survey via department email lists and social media, inviting the participants to share their `computer-usage patterns and behavior change needs'. There were no explicit inclusion or exclusion criteria for the participants and the participants did not receive any compensation for the survey.

\textsc{Participants} We had 68 participants (35 males, 33 females; $\mu$ = 32.9 years, $\sigma$ = 14.8 years; 28 students, 39 full-time workers, 1 retired) from 9 countries -- 35 from the United States (6 different states), 27 from Malaysia, 24 from the United Kingdom, 4 from Pakistan, 2 from Canada, 2 from India, 1 from France, 2 from Singapore, and 1 from Germany.

\textsc{Survey Questions:} We created our own survey since there was no preexisting survey to investigate our three research questions. We started with Likert scale questions to minimally survey the overall computer usage patterns (Q1-2) and broad problems categories (Q3) of the participants. We did not include full standardized computer usage surveys like CUS to keep our survey short. We then included open-ended questions (Q4-9) to survey the diverse and detailed experiences of our participants for each of our research questions -- \textbf{RQ1:} Computer-usage behavior change needs (Q4) and specific desired changes (Q5);
\textbf{RQ2:} Currently-used computer-usage behavior change applications and if and why they work or do not work (Q6), and further-desired support (Q7);
\textbf{RQ3:} User experiences with helpful computer breaks and why they are helpful (Q8), and similarly for unhelpful breaks (Q9). 
We iteratively developed the survey questions via peer review and expert review (5 experts and 5 peers) to ensure the validity and reliability of our questions. We also did 5 pilot surveys to further check validity and reliability.
All survey questions are in Table \ref{tab:survey-questions}. 

\textsc{Data Analysis:} For each of the open-ended questions (Q4-9), three researchers independently coded the responses and then collectively performed a thematic analysis of the responses. We performed inductive analysis and the 3 researchers iterated on the codes, themes, and categories for each question before finalizing them. We share the coded responses, themes, and also the top 50 words in each of the responses (excluding words repeated from the question). 

\begin{table}[]
  \centering
\caption{Survey questions about computer-usage patterns, behavior change needs, and helpful/unhelpful breaks}
\label{tab:survey-questions}
\begin{tabular}{ll}
\hline
\textbf{Q} & \textbf{Survey Questions} \\ \hline 
\multicolumn{2}{c}{\textbf{Computer-Usage Patterns and Overall Problems}}
\\
1 & How much time do you spend using your computer daily?  5 options: 0-2, 2-5, 5-10, 10-15, >15 hours \\
2 & How much do you use your computer/internet for the following (daily)? 4 categories and 5 options each:
\\ & Work/Learn, Social Network, Fun/Relax, Miscellaneous (Options: 0-2, 2-5, 5-10, 10-15, >15 hrs) \\
3 & How often do you experience the following side-effects due to your computer/internet usage? \\ & 5 categories: Poor time management/distraction, Emotional stress, Physical discomfort, Social problems, 
\\
& Financial problems; 5 single-select options for each category: Never, Rarely, Sometimes, Often, Always
\\
\multicolumn{2}{c}{\textbf{RQ1. Computer-related Behavior Change Needs and Specific Desired Changes}}
\\
4 & Is there anything you would like to change about your computer usage? Why and how? \\
5 & Is there something you'd like to spend less time on or more time on? Please be specific, e.g., give examples.\\
\multicolumn{2}{c}{\textbf{RQ2. Currently-used Computer-usage Behavior Change Tools and Further Needs}}
\\
6 & Is there anything you use or have used to manage your computer usage? Does it work? Why or why not?\\
7 & Is there something you would like to help you manage your computer usage? \\
\multicolumn{2}{c}{\textbf{RQ3. Computer-related Helpful and Unhelpful Breaks}}
\\
8 & Think of an example of a helpful computer break you took. (a) What activity did you do and how did it come\\
& about? (b) How long was the break? (c) Why was it helpful? \\
9 & Think of an example of an unhelpful computer break you took. (a) What was the break and how did it come
\\
& about? (b) How long was the break? (c) Why was it unhelpful? 
\\
\hline
\end{tabular}
\end{table}

\section{Results}

We summarize below the results from our survey below for each of the research questions: RQ1. Computer-related Problems and Behavior Change Needs (Q4-5); RQ2. Currently-used Computer-usage Behavior Change Tools and Further Needs (Q6-7); RQ3. Computer-related Helpful and Unhelpful Breaks (Q8-9). Also, we summarize the computer-usage patterns (Q1-Q2) and overall problems (Q3) below.

\begin{figure}[]
  \includegraphics[width=\textwidth]{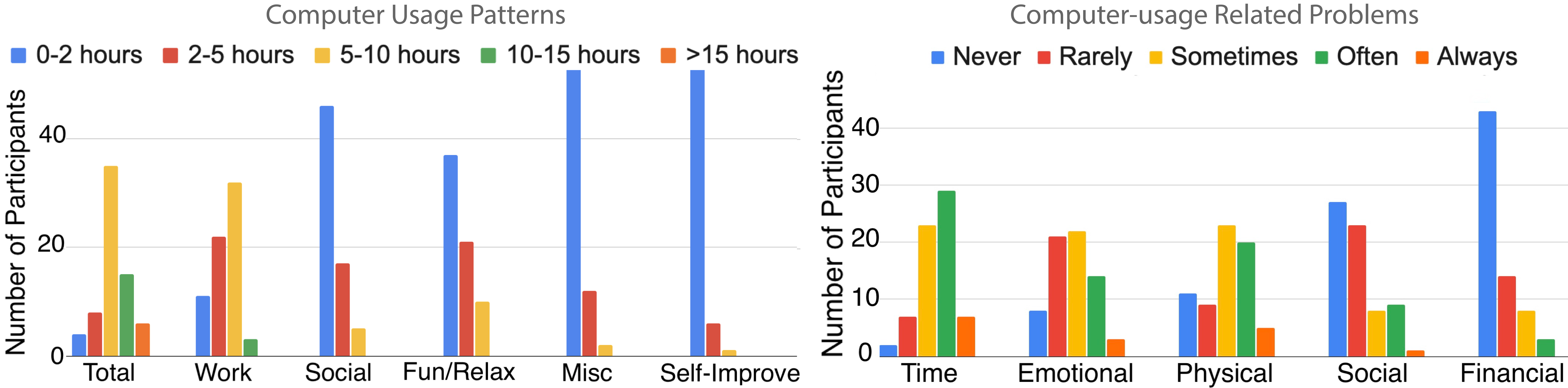}
  \caption{Likert scale responses. Left: Computer Usage Patterns (Q1, 2). Right: Computer-usage Problems (Q3)}
  \label{fig:q3}
  \vspace{-0.5em}
\end{figure}

\textbf{Q1, 2. Computer Usage Duration:} Most participants reported `5-10 hours' of total computer usage.
For `Work/ Learning', the most common duration was `5-10 hours', and for `Social Networking', `Fun/Relaxation', and `Miscellaneous', the most common duration was `0-2 hours' each.
The detailed results are shown in Figure \ref{fig:q3}.

\textbf{Q3. Computer-Related Problems: }
Most participants `Often' experienced `Time management' problems, `Sometimes' experienced `Emotional stress' and `Physical discomfort', and `Never' experienced `Social problems' and 'Financial problems' (though `Sometimes' was a close second for `Social problems').
The detailed results are shown in Figure \ref{fig:q3}.

\subsection{RQ1. Computer-related Problems and Behavior Change Needs (Q4-5)} 

\begin{figure}[]
  \centering
  \includegraphics[width=0.74\textwidth]{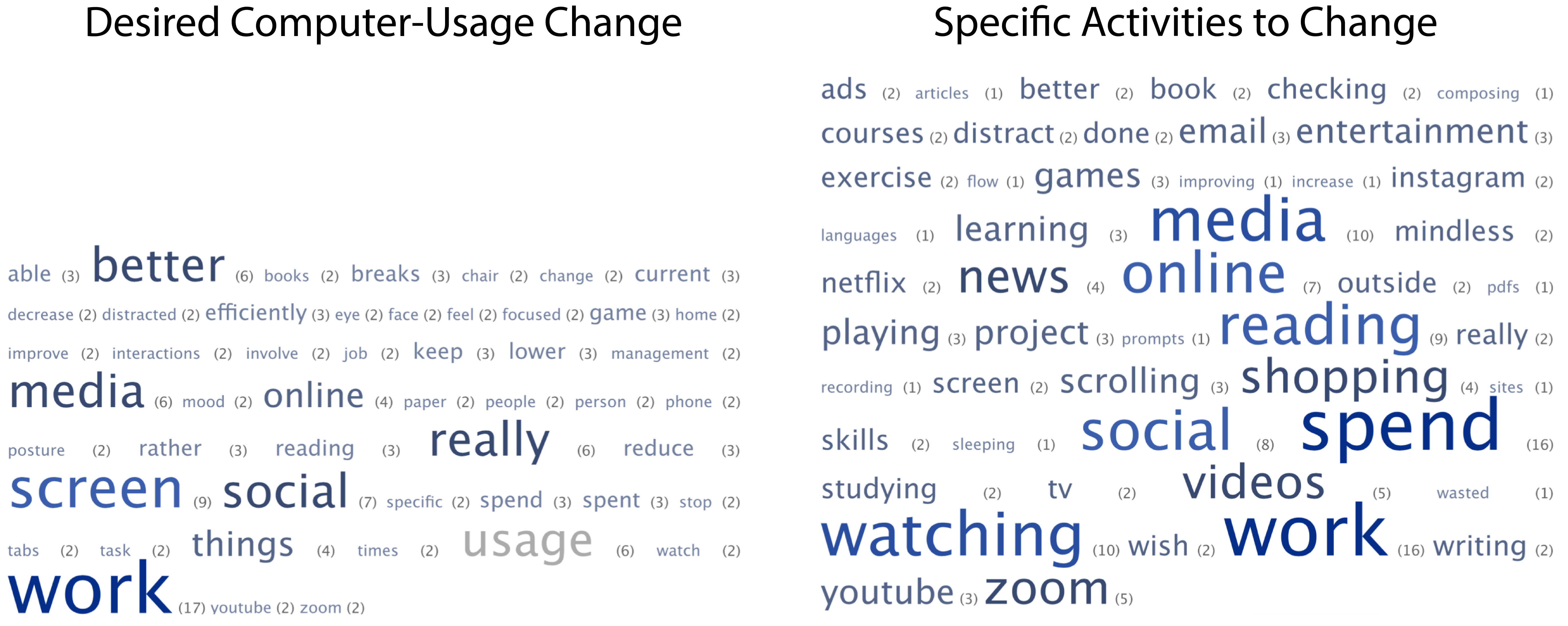}
  \caption{Top 50 words in open-ended responses. Left: Desired computer-usage change (Q4). Right: Specific activities change (Q5)}
  \label{fig:w45}
\end{figure}

\textbf{Q4. Desired Changes: }
45 participants responded with specific desired changes (Q4), and we coded the responses into five categories -- Reduce time drain (20), Reduce social media (10), Reduce physical side-effects (9), Reduce entertainment (3), and Improve device set-up (3).
\textbf{Q5. Desired Activities and Reasons:} The participants mentioned several different specific desired activities (Q5), and we grouped the responses into 10 categories reflecting the underlying reasons for change -- `Better time management' (12), `Feel emotionally better' (8), `Reduce checking phone' (6), `Better posture/reduce eye strain '(6), `Less addictive scrolling' (4), etc.
For both Q4 and Q5, time management, emotional well-being, and physical health were common themes. Figure \ref{fig:sankey} shows the coded responses for desired changes (Q4) and reasons for activity changes (Q5).
Figure \ref{fig:w45} shows the top 50 words in the responses to Q4 (left) and Q5 (right).

\subsection{RQ2. Currently-used Computer-usage Behavior Change Tools and Further Needs (Q6-7)}

\begin{figure}[]
  \centering
  \includegraphics[width=0.495\textwidth]{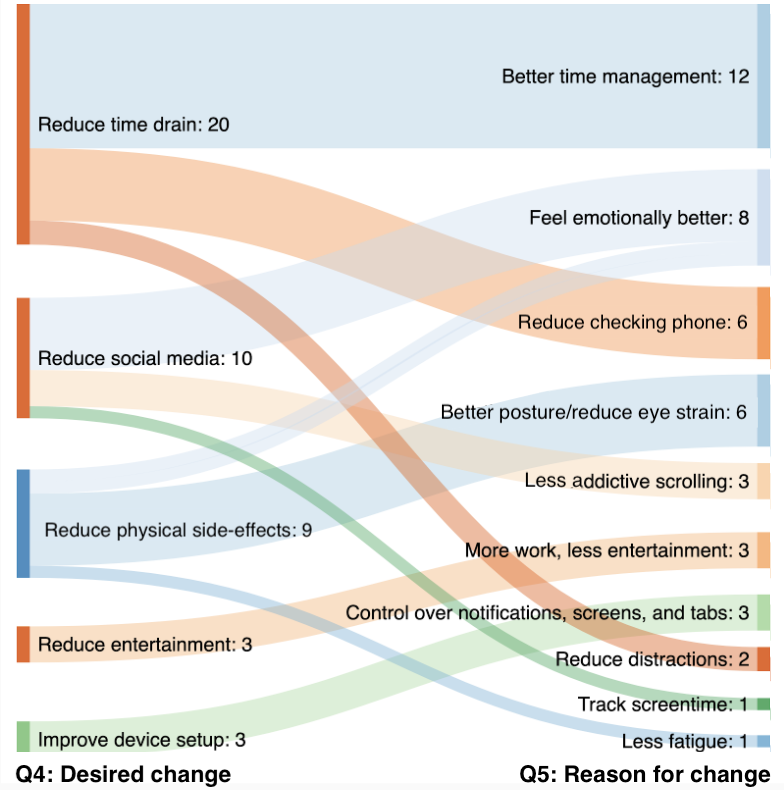}
  \includegraphics[width=0.495\textwidth]{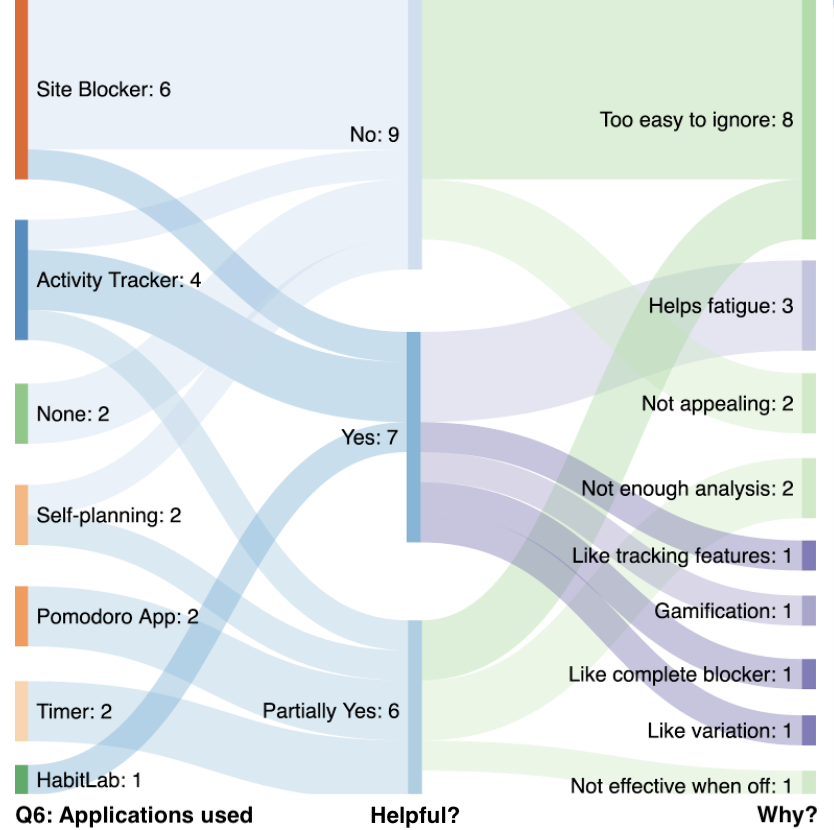}
  \caption{Coded responses for open-ended questions (Q4-6). Left: Desired computer-usage behavior change (Q4) and specific desired activities (Q5). Right: Currently used support for computer-usage behavior change and why they do or do not work (Q6)}
  \label{fig:sankey}
  \vspace{-0.5em}
\end{figure}

\textbf{Q6. Tools Used and their Efficacy: }
We received 22 responses about the tools used by the participants for computer-related behavior change. Site blockers were the most commonly used types of tools, followed by activity trackers, and then equally by self-planning tools, Pomodoro-technique apps, and timers. However, most of the applications used by participants were not helpful, mostly because they were `Easy to ignore' or `Not appealing'. Reasons why some applications were helpful included `helps fatigue', `like tracking features', `gamification', `like complete blocker', and `like variation'. Figure \ref{fig:sankey} (right) shows the categories of tools used, their efficacy, and the reasons behind their efficacy.

\begin{figure}[]
  \centering
  \includegraphics[width=\textwidth]{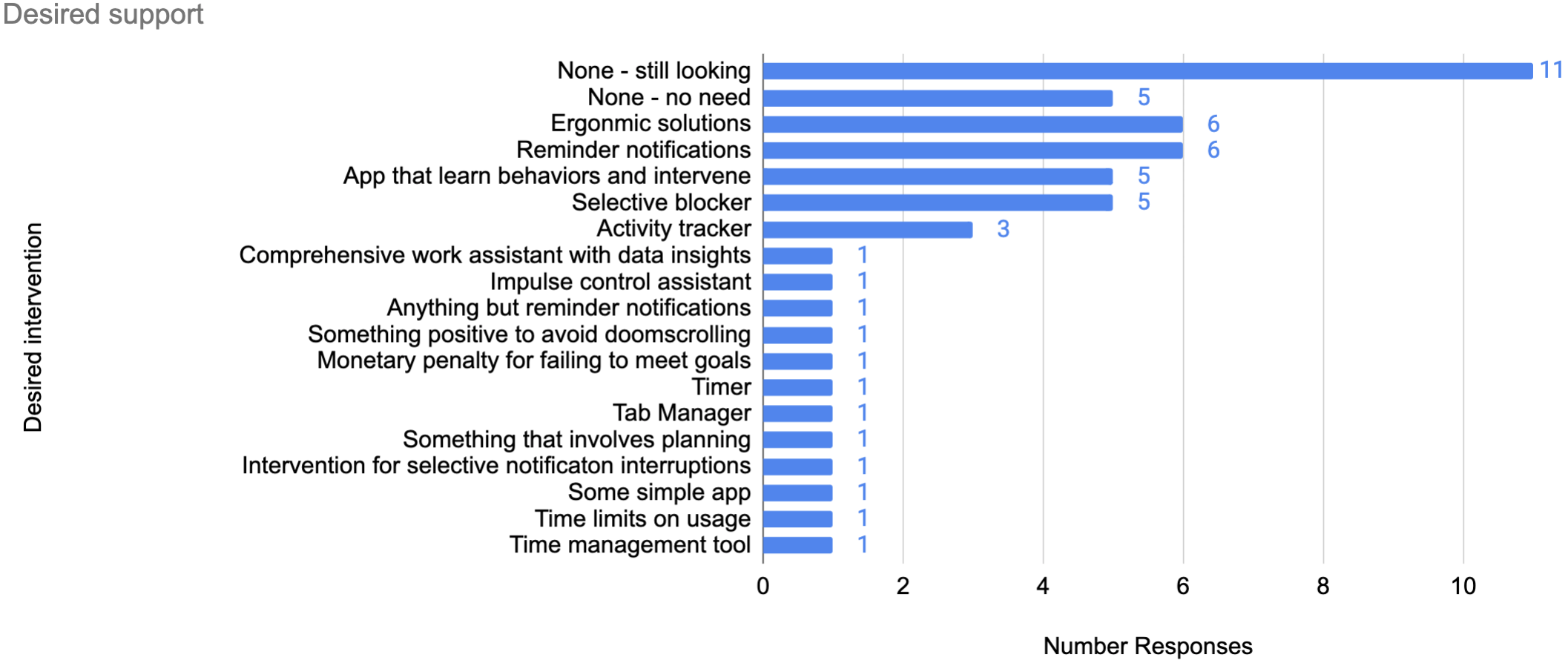}
  \caption{Open-ended responses for desired computer-usage behavior change support (Q7)}
  \label{fig:further_desired_support}
  \vspace{-0.5em}
\end{figure}

\textbf{Q7. Further Support Desired: }
53 participants responded - 11 did not know what they wanted, 5 did not need any, and rest had diverse suggestions. We created 17 categories for the responses, e.g., ergonomics solutions (6), reminders notifications (6), learned behavior and interventions (5), selective blockers (5), etc. The coded responses are in Figure \ref{fig:further_desired_support}.

\begin{figure}[]
  \centering
\includegraphics[width=0.74\textwidth]{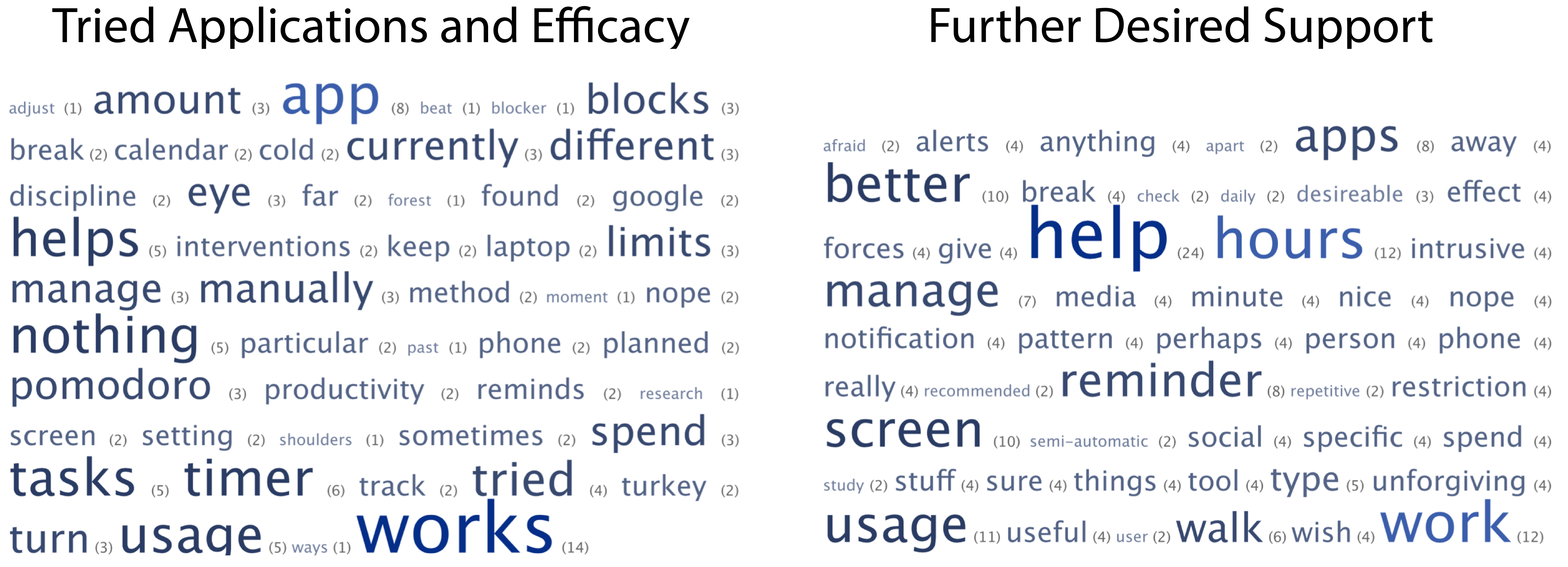}
  \caption{Top 50 words in open-ended responses. Left: Tried applications and their efficacy (Q6). Right: Further desired support (Q7)}
  \label{fig:w67}
  \vspace{-0.5em}
\end{figure}

Figure \ref{fig:w67}  shows the top 50 words in the responses to Q6 (left) and Q7 (right).

\subsection{RQ3. Computer-related Helpful and Unhelpful Breaks (Q8-9)}

\begin{figure}[]
  \centering
  \includegraphics[width=0.495\textwidth]{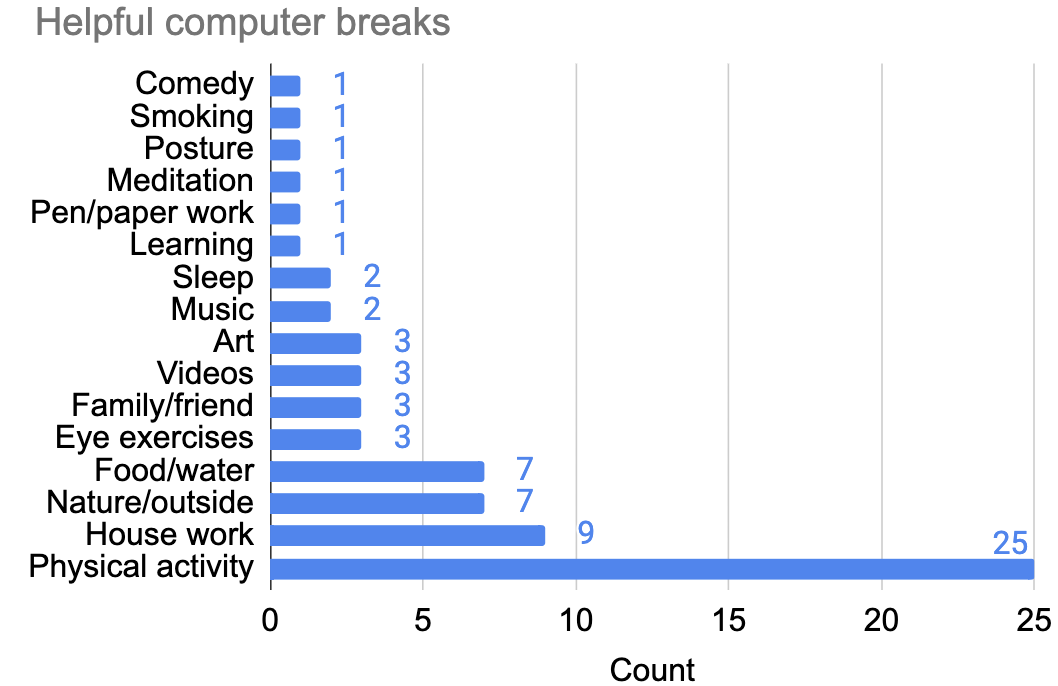}
  \includegraphics[width=0.495\textwidth]{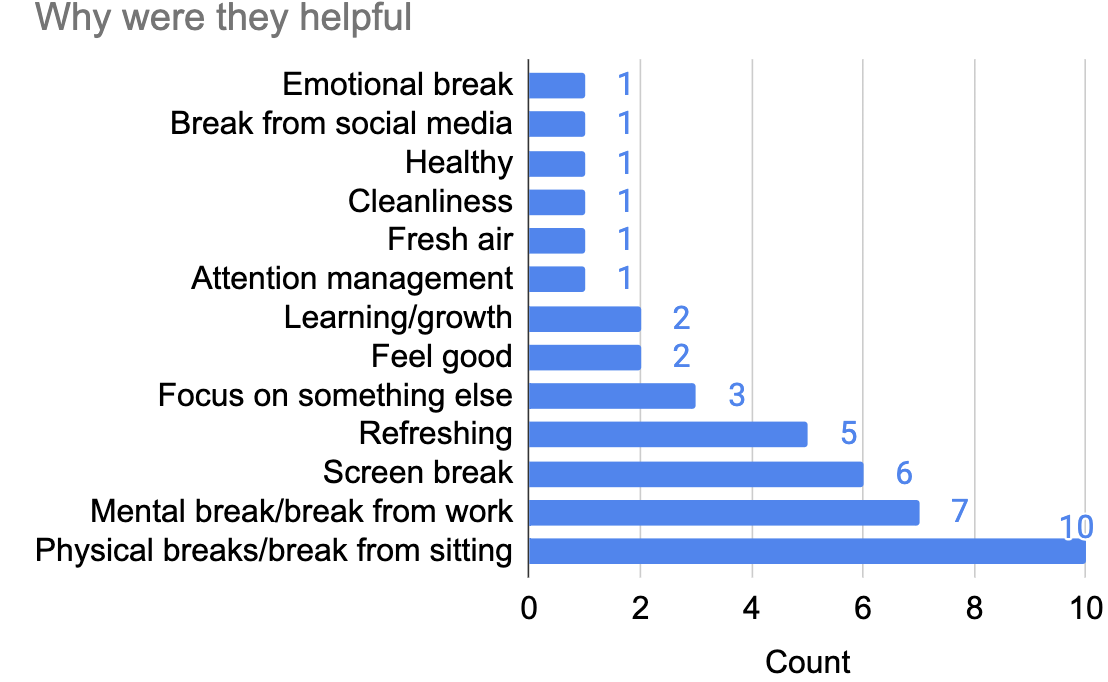}
  \caption{Q8. Coded open-ended responses for helpful computer-related breaks (left) and reasons why they were helpful (right)}
  \label{fig:helpful_breaks}
  \vspace{-0.5em}
\end{figure}

\textbf{Q8. Helpful Breaks and Reasons:}
We divided the responses into 16 categories. 
Some participants mentioned multiple breaks or one break that fell into multiple categories, e.g., housework also involved physical activity. 
The most common category of helpful breaks was physical activity (25), followed by housework (9), time in nature/outside (7), food/water breaks (7), eye exercises (3), family/friends time (3), watching videos (3), doing art (3), etc. 
Overall, most of the helpful breaks were away from the computer and the breaks were diverse, e.g., `Nature/outside' was gardening for some and watching the sunset for others. 
The duration of helpful breaks was diverse: <5 min (5), 5-15 min (8), 15-30 min (14), 30-60 min (9), 1 hour-1 day (4), >1 day (2).
41 participants mentioned the reasons for `helpful' breaks, and we divided the responses into 13 categories -- Physical breaks/ break from sitting (10), Mental break/ break from work (7), Screen break (6), Refreshing (5), Focus on something else (3), Feel good (2), Learning/ growth (2), etc. The results are in Figure \ref{fig:helpful_breaks}.

\begin{figure}[]
  \centering
  \includegraphics[width=0.515\textwidth]{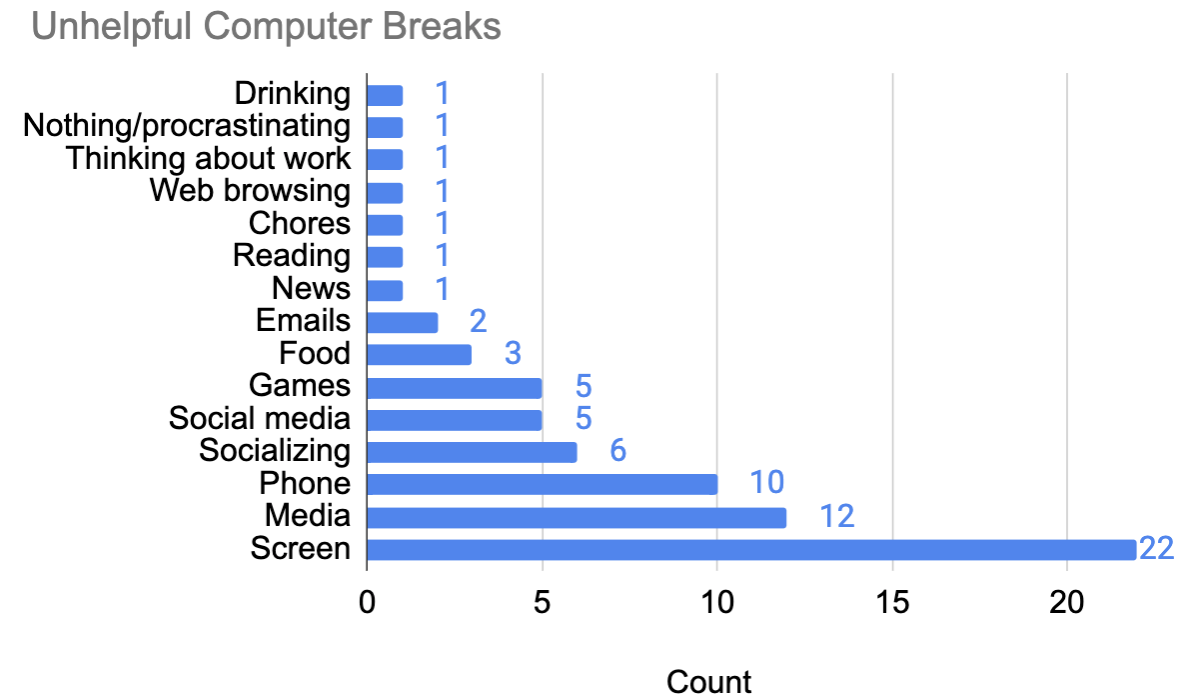}
  \hspace{0.7 cm}
  \includegraphics[width=0.42\textwidth]{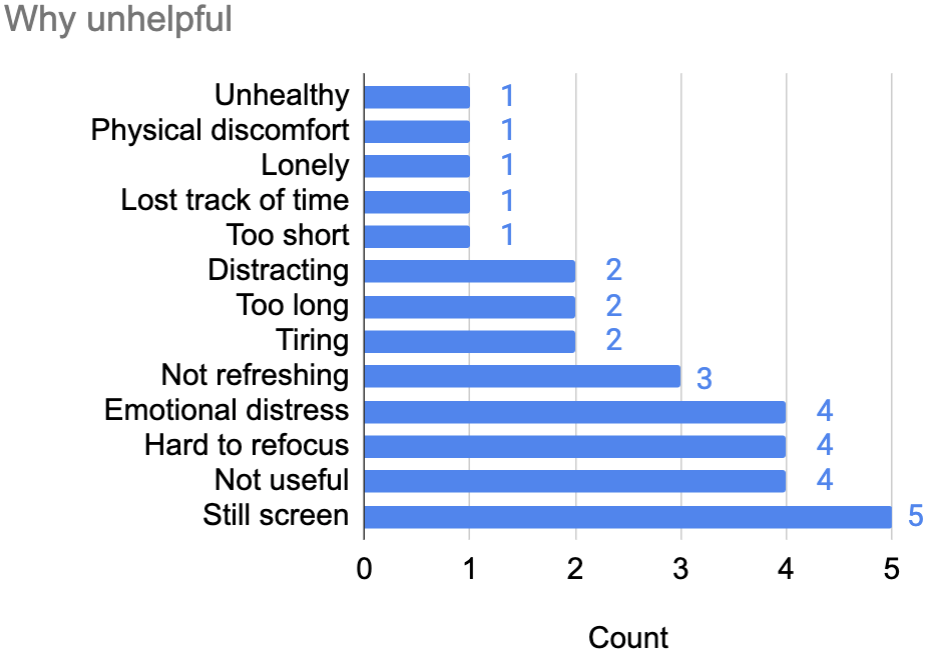}
  \caption{Q9. Open-ended responses for unhelpful computer-related breaks (left) and reasons why they were unhelpful (right)}
  \label{fig:unhelpful_breaks}
  \vspace{-0.5em}
\end{figure}

\textbf{Q9. Unhelpful Breaks and Reasons:}
We divided the responses into 15 categories, which had some overlaps, e.g., `screen' category was also connected to `phone', `media' and `web browsing', but because some participants explicitly mentioned only one, we characterized the responses to closest mentioned category.
Most of the unhelpful breaks involved phone or computer screens, e.g., games (5), social media (5), web browsing (1), and videos (12), or long social interactions (6).
7 participants mentioned the duration of unhelpful breaks, and 6 of them reported more than 30-minute-long breaks: 15-20 min (1), 30-60 min (5), 1-3 hours (1).
Finally, 31 participants mentioned why they found the break unhelpful, and we divided the responses into 13 categories -- `still screen' (5), `not useful' (4), `hard to refocus' (4), `emotional distress' (4), `not refreshing' (3), `tiring' (2), `too long' (2), `distracting' (2), etc. The results are in Figure \ref{fig:unhelpful_breaks}.

\begin{figure}[]
  \centering
  \includegraphics[width=0.75\textwidth]{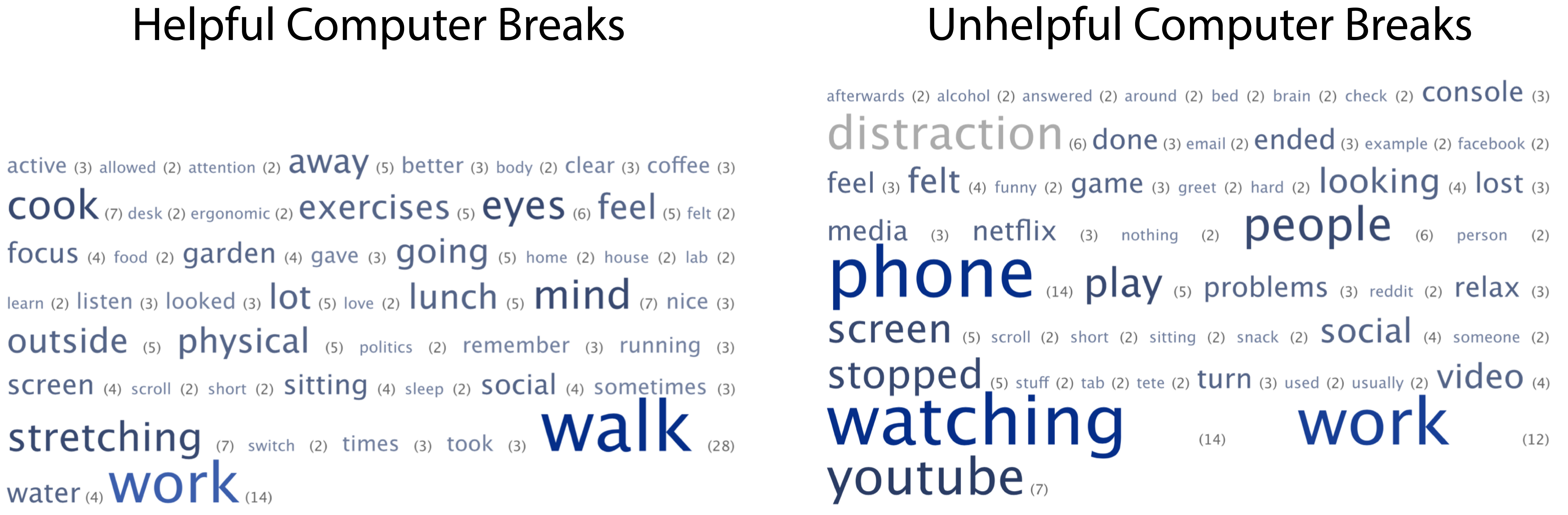}

  \caption{Top 50 words in responses to Q8: Helpful computer-related breaks (left) and Q9: Unhelpful computer-related breaks (right)}
  \label{fig:w89}
  \vspace{-0.5em}
\end{figure}

Figure \ref{fig:w89} shows the raw top 50 words used in survey responses to Q8 (left) and Q9 (right).

\section{Discussion and Recommendations}

Previous computer-related behavior change work has focused on specific goals, e.g., improving productivity, focus and physical activity. We conducted a study to identify the overall user needs. We discuss the key findings, limitations, and recommendations of our computer-usage behavior change needs user survey below.

\subsection{Findings}
Most people spent between 5-10 hours on their computer (Q1, Q2), and time management, emotional problems, and physical discomfort are key concerns for people with respect to their computer usage (Q3). We had three key findings from our survey. \textbf{First}, the participants wanted to reduce time drain, social media usage, entertainment, and physical side-effects (Q4) to better manage their time, emotions, and physical health, and also do less addictive scrolling and `phone checking' (Q5).
Users needs were, thus, have diverse and intertwined needs as they want to reduce their time drain, social media usage, and physical discomfort (Q4) to better manage their time and physical and emotional health (Q5). \textbf{Second}, people use site blockers, time management, and self-tracking applications, but many do not work well, especially because they are easy to ignore and are not designed for different user needs in different contexts (Q6). Thus, users want personalized interventions (e.g., something positive) in personalized contexts (e.g., selective blockers) and with personalized self-tracking insights (Q7). \textbf{Third}, away-from-screen breaks were usually helpful, especially when they were under 1 hour and involved physical or mental breaks (Q8), whereas on-screen breaks were not helpful and even tended to leave people exhausted and demotivated to work, especially when they were longer than 1 hour (Q9).

\subsection{Limitations}
We highlight three limitations of our work. 
\textbf{First}, we conducted our survey during the COVID 2019 pandemic and the computer-usage behavior change needs may be different than `normal' times since most work was done virtually. However, given the increasing reliance on technology, our survey highlights the current and future needs for computer-usage behavior change.
\textbf{Second}, our survey shows diverse participant needs and responses and does not explore one specific area, e.g., overcoming distractions or sedentary behavior. However, it is important to highlight and harness the diverse user needs and experiences to support real-life user needs.
\textbf{Third}, we focus on user needs, but the user may not know what is best for them, e.g., user's perceived efficacy may be different from actual efficacy (Q6), and user reports on helpful/unhelpful breaks (Q8, 9) may be biased memories. Thus, it is important to be mindful of user needs and experiences, but then also test them via in situ and longitudinal studies.
Overall, our survey is the first of its kind and further surveys and studies may be needed to validate our findings in different settings and with different users. 

\subsection{Recommendations}
We have three recommendations.
\textbf{First}, even though computer-related behavior change has focused on disjoint goals like productivity, focus, and physical activity, computer-usage behavior change needs are diverse and interconnected -- reducing distractions is connected to better time management and productivity, and even physical activity breaks are helpful for ``better'' computer usage (Q8). The three broad categories of user needs are time management, emotional wellbeing, and physical health, and it may help to provide holistic and personalizable support for user's diverse and interconnected user needs.
\textbf{Second}, computer-usage behavior change support can be easy-to-ignore and users want personalized and closed-loop support. We recommend closed-loop behavior change support using reinforcement learning to monitor each user and provide better personalized and context-aware support.
\textbf{Third,} off-the-screen breaks are more helpful than on-the-screen breaks. Instead of only restricting on-the-screen breaks, especially since restrictions can be stressful \cite{mark2017blocking}, it might help to replace on-the-screen draining breaks with off-the-screen refreshing breaks.
Also, compulsive technology use is a problem \cite{clements2018compulsive} and technology use has a cognitive cost \cite{kang2019reach}. 
Thus, we recommend encouraging off-the-screen breaks to curb on-the-screen distractions and boost on-the-screen productivity. 

\section{Conclusion}
Previous research on computer-related behavior change has focused on goals like productivity, focus, or physical activity, or on evaluating what users consider as computer breaks and what are helpful breaks for productivity and focus.
We conducted a need-finding survey to investigate three key research questions with respect to people's computer-usage change: i. what do people want to change and why; ii. what do people currently use or have used and what is further desired; iii. what are helpful and unhelpful breaks and why.

Our findings show that user needs for computer-related behavior change are diverse and interconnected, e.g., the need for better time management is connected to better productivity and focus and even better physical activity as physical movement enables ``better'' overall computer usage experience.
Many computer-usage-related behavior change applications do not work for the users as they are easy-to-ignore and also not personalized for users.
Finally, off-the-screen breaks are, in general, helpful whereas on-the-screen breaks are not helpful.

Thus, user needs are interconnected and users need personalized, holistic, and closed-loop support. One way to offer personalized and closed-loop support is to use reinforcement learning to learn the best context-aware interventions for each user. Also, off-the-computer breaks may be more helpful than on-the-computer breaks.
We believe that our findings will inform the design of future behavior change applications for computer usage.

\bibliographystyle{ACM-Reference-Format}
\bibliography{sample-base}


\begin{thebibliography}{32}


\ifx \showCODEN    \undefined \def \showCODEN     #1{\unskip}     \fi
\ifx \showDOI      \undefined \def \showDOI       #1{#1}\fi
\ifx \showISBNx    \undefined \def \showISBNx     #1{\unskip}     \fi
\ifx \showISBNxiii \undefined \def \showISBNxiii  #1{\unskip}     \fi
\ifx \showISSN     \undefined \def \showISSN      #1{\unskip}     \fi
\ifx \showLCCN     \undefined \def \showLCCN      #1{\unskip}     \fi
\ifx \shownote     \undefined \def \shownote      #1{#1}          \fi
\ifx \showarticletitle \undefined \def \showarticletitle #1{#1}   \fi
\ifx \showURL      \undefined \def \showURL       {\relax}        \fi
\providecommand\bibfield[2]{#2}
\providecommand\bibinfo[2]{#2}
\providecommand\natexlab[1]{#1}
\providecommand\showeprint[2][]{arXiv:#2}

\bibitem[\protect\citeauthoryear{Barata, Nicolau, and Gonçalves}{Barata
  et~al\mbox{.}}{2012}]%
        {barata_appinsight_2012}
\bibfield{author}{\bibinfo{person}{Gabriel Barata}, \bibinfo{person}{Hugo
  Nicolau}, {and} \bibinfo{person}{Daniel Gonçalves}.}
  \bibinfo{year}{2012}\natexlab{}.
\newblock \showarticletitle{{AppInsight}: what have {I} been doing?}. In
  \bibinfo{booktitle}{\emph{Proceedings of the {International} {Working}
  {Conference} on {Advanced} {Visual} {Interfaces}}}
  \emph{(\bibinfo{series}{{AVI} '12})}. \bibinfo{publisher}{Association for
  Computing Machinery}, \bibinfo{address}{New York, NY, USA},
  \bibinfo{pages}{465--472}.
\newblock
\showISBNx{978-1-4503-1287-5}
\urldef\tempurl%
\url{https://doi.org/10.1145/2254556.2254645}
\showDOI{\tempurl}


\bibitem[\protect\citeauthoryear{Borghouts, Brumby, and Cox}{Borghouts
  et~al\mbox{.}}{2020}]%
        {borghouts_timetofocus_2020}
\bibfield{author}{\bibinfo{person}{Judith Borghouts},
  \bibinfo{person}{Duncan~P. Brumby}, {and} \bibinfo{person}{Anna~L. Cox}.}
  \bibinfo{year}{2020}\natexlab{}.
\newblock \showarticletitle{{TimeToFocus}: {Feedback} on {Interruption}
  {Durations} {Discourages} {Distractions} and {Shortens} {Interruptions}}.
\newblock \bibinfo{journal}{\emph{ACM Transactions on Computer-Human
  Interaction}} \bibinfo{volume}{27}, \bibinfo{number}{5} (\bibinfo{date}{Aug.}
  \bibinfo{year}{2020}), \bibinfo{pages}{32:1--32:31}.
\newblock
\showISSN{1073-0516}
\urldef\tempurl%
\url{https://doi.org/10.1145/3396044}
\showDOI{\tempurl}


\bibitem[\protect\citeauthoryear{Cambo, Avrahami, and Lee}{Cambo
  et~al\mbox{.}}{2017}]%
        {cambo2017breaksense}
\bibfield{author}{\bibinfo{person}{Scott~A Cambo}, \bibinfo{person}{Daniel
  Avrahami}, {and} \bibinfo{person}{Matthew~L Lee}.}
  \bibinfo{year}{2017}\natexlab{}.
\newblock \showarticletitle{BreakSense: Combining physiological and location
  sensing to promote mobility during work-breaks}. In
  \bibinfo{booktitle}{\emph{Proceedings of the 2017 CHI Conference on Human
  Factors in Computing Systems}}. \bibinfo{pages}{3595--3607}.
\newblock


\bibitem[\protect\citeauthoryear{Cavdar, Taskaya-Temizel, Musolesi, and
  Tino}{Cavdar et~al\mbox{.}}{2020}]%
        {cavdar_multi-perspective_2020}
\bibfield{author}{\bibinfo{person}{Seyma~Kucukozer Cavdar},
  \bibinfo{person}{Tugba Taskaya-Temizel}, \bibinfo{person}{Mirco Musolesi},
  {and} \bibinfo{person}{Peter Tino}.} \bibinfo{year}{2020}\natexlab{}.
\newblock \showarticletitle{A {Multi}-perspective {Analysis} of {Social}
  {Context} and {Personal} {Factors} in {Office} {Settings} for the {Design} of
  an {Effective} {Mobile} {Notification} {System}}.
\newblock \bibinfo{journal}{\emph{Proceedings of the ACM on Interactive,
  Mobile, Wearable and Ubiquitous Technologies}} \bibinfo{volume}{4},
  \bibinfo{number}{1} (\bibinfo{date}{March} \bibinfo{year}{2020}),
  \bibinfo{pages}{15:1--15:38}.
\newblock
\urldef\tempurl%
\url{https://doi.org/10.1145/3381000}
\showDOI{\tempurl}


\bibitem[\protect\citeauthoryear{Choi, Park, Kim, Lim, and Lee}{Choi
  et~al\mbox{.}}{2019}]%
        {choi_multi-stage_2019}
\bibfield{author}{\bibinfo{person}{Woohyeok Choi}, \bibinfo{person}{Sangkeun
  Park}, \bibinfo{person}{Duyeon Kim}, \bibinfo{person}{Youn-kyung Lim}, {and}
  \bibinfo{person}{Uichin Lee}.} \bibinfo{year}{2019}\natexlab{}.
\newblock \showarticletitle{Multi-{Stage} {Receptivity} {Model} for {Mobile}
  {Just}-{In}-{Time} {Health} {Intervention}}.
\newblock \bibinfo{journal}{\emph{Proceedings of the ACM on Interactive,
  Mobile, Wearable and Ubiquitous Technologies}} \bibinfo{volume}{3},
  \bibinfo{number}{2} (\bibinfo{date}{June} \bibinfo{year}{2019}),
  \bibinfo{pages}{39:1--39:26}.
\newblock
\urldef\tempurl%
\url{https://doi.org/10.1145/3328910}
\showDOI{\tempurl}


\bibitem[\protect\citeauthoryear{Clement}{Clement}{2020}]%
        {clement2020coronavirus}
\bibfield{author}{\bibinfo{person}{J Clement}.}
  \bibinfo{year}{2020}\natexlab{}.
\newblock \showarticletitle{Coronavirus: Impact on Online Usage in the
  US-Statistics \& Facts}.
\newblock \bibinfo{journal}{\emph{Hamburg: Statista}} (\bibinfo{year}{2020}).
\newblock


\bibitem[\protect\citeauthoryear{Clements and Boyle}{Clements and
  Boyle}{2018}]%
        {clements2018compulsive}
\bibfield{author}{\bibinfo{person}{Jeffrey~A Clements} {and}
  \bibinfo{person}{Randall Boyle}.} \bibinfo{year}{2018}\natexlab{}.
\newblock \showarticletitle{Compulsive technology use: Compulsive use of mobile
  applications}.
\newblock \bibinfo{journal}{\emph{Computers in Human behavior}}
  \bibinfo{volume}{87} (\bibinfo{year}{2018}), \bibinfo{pages}{34--48}.
\newblock


\bibitem[\protect\citeauthoryear{Di~Lascio, Gashi, Hidalgo, Nale, Debus, and
  Santini}{Di~Lascio et~al\mbox{.}}{2020}]%
        {di_lascio_multi-sensor_2020}
\bibfield{author}{\bibinfo{person}{Elena Di~Lascio}, \bibinfo{person}{Shkurta
  Gashi}, \bibinfo{person}{Juan~Sebastian Hidalgo}, \bibinfo{person}{Beatrice
  Nale}, \bibinfo{person}{Maike~E. Debus}, {and} \bibinfo{person}{Silvia
  Santini}.} \bibinfo{year}{2020}\natexlab{}.
\newblock \showarticletitle{A {Multi}-{Sensor} {Approach} to {Automatically}
  {Recognize} {Breaks} and {Work} {Activities} of {Knowledge} {Workers} in
  {Academia}}.
\newblock \bibinfo{journal}{\emph{Proceedings of the ACM on Interactive,
  Mobile, Wearable and Ubiquitous Technologies}} \bibinfo{volume}{4},
  \bibinfo{number}{3} (\bibinfo{date}{Sept.} \bibinfo{year}{2020}),
  \bibinfo{pages}{78:1--78:20}.
\newblock
\urldef\tempurl%
\url{https://doi.org/10.1145/3411821}
\showDOI{\tempurl}


\bibitem[\protect\citeauthoryear{Epstein, Avrahami, and Biehl}{Epstein
  et~al\mbox{.}}{2016}]%
        {epstein2016taking}
\bibfield{author}{\bibinfo{person}{Daniel~A Epstein}, \bibinfo{person}{Daniel
  Avrahami}, {and} \bibinfo{person}{Jacob~T Biehl}.}
  \bibinfo{year}{2016}\natexlab{}.
\newblock \showarticletitle{Taking 5: Work-breaks, productivity, and
  opportunities for personal informatics for knowledge workers}. In
  \bibinfo{booktitle}{\emph{Proceedings of the 2016 CHI Conference on Human
  Factors in Computing Systems}}. \bibinfo{pages}{673--684}.
\newblock


\bibitem[\protect\citeauthoryear{Guillou, Chow, Fritz, and McGrenere}{Guillou
  et~al\mbox{.}}{2020}]%
        {guillou_is_2020}
\bibfield{author}{\bibinfo{person}{Hayley Guillou}, \bibinfo{person}{Kevin
  Chow}, \bibinfo{person}{Thomas Fritz}, {and} \bibinfo{person}{Joanna
  McGrenere}.} \bibinfo{year}{2020}\natexlab{}.
\newblock \showarticletitle{Is {Your} {Time} {Well} {Spent}? {Reflecting} on
  {Knowledge} {Work} {More} {Holistically}}. In
  \bibinfo{booktitle}{\emph{Proceedings of the 2020 {CHI} {Conference} on
  {Human} {Factors} in {Computing} {Systems}}} \emph{(\bibinfo{series}{{CHI}
  '20})}. \bibinfo{publisher}{Association for Computing Machinery},
  \bibinfo{address}{New York, NY, USA}, \bibinfo{pages}{1--9}.
\newblock
\showISBNx{978-1-4503-6708-0}
\urldef\tempurl%
\url{https://doi.org/10.1145/3313831.3376586}
\showDOI{\tempurl}


\bibitem[\protect\citeauthoryear{Healy}{Healy}{1999}]%
        {healy1999failure}
\bibfield{author}{\bibinfo{person}{Jane~M Healy}.}
  \bibinfo{year}{1999}\natexlab{}.
\newblock \bibinfo{booktitle}{\emph{Failure to connect: How computers affect
  our children's minds--for better and worse}}.
\newblock \bibinfo{publisher}{Simon and Schuster}.
\newblock


\bibitem[\protect\citeauthoryear{Hu and Lee}{Hu and Lee}{2019}]%
        {hu_screentrack_2019}
\bibfield{author}{\bibinfo{person}{Donghan Hu} {and} \bibinfo{person}{Sang~Won
  Lee}.} \bibinfo{year}{2019}\natexlab{}.
\newblock \showarticletitle{{ScreenTrack}: {Using} {Visual} {History} for
  {Self}-tracking {Computer} {Activities} and {Retrieving} {Working}
  {Context}}. In \bibinfo{booktitle}{\emph{The {Adjunct} {Publication} of the
  32nd {Annual} {ACM} {Symposium} on {User} {Interface} {Software} and
  {Technology}}} \emph{(\bibinfo{series}{{UIST} '19})}.
  \bibinfo{publisher}{Association for Computing Machinery},
  \bibinfo{address}{New York, NY, USA}, \bibinfo{pages}{44--46}.
\newblock
\showISBNx{978-1-4503-6817-9}
\urldef\tempurl%
\url{https://doi.org/10.1145/3332167.3357110}
\showDOI{\tempurl}


\bibitem[\protect\citeauthoryear{Kang and Kurtzberg}{Kang and
  Kurtzberg}{2019}]%
        {kang2019reach}
\bibfield{author}{\bibinfo{person}{Sanghoon Kang} {and}
  \bibinfo{person}{Terri~R Kurtzberg}.} \bibinfo{year}{2019}\natexlab{}.
\newblock \showarticletitle{Reach for your cell phone at your own risk: The
  cognitive costs of media choice for breaks}.
\newblock \bibinfo{journal}{\emph{Journal of behavioral addictions}}
  \bibinfo{volume}{8}, \bibinfo{number}{3} (\bibinfo{year}{2019}),
  \bibinfo{pages}{395--403}.
\newblock


\bibitem[\protect\citeauthoryear{Kaur, Williams, McDuff, Czerwinski, Teevan,
  and Iqbal}{Kaur et~al\mbox{.}}{2020}]%
        {kaur_optimizing_2020}
\bibfield{author}{\bibinfo{person}{Harmanpreet Kaur}, \bibinfo{person}{Alex~C.
  Williams}, \bibinfo{person}{Daniel McDuff}, \bibinfo{person}{Mary
  Czerwinski}, \bibinfo{person}{Jaime Teevan}, {and} \bibinfo{person}{Shamsi~T.
  Iqbal}.} \bibinfo{year}{2020}\natexlab{}.
\newblock \showarticletitle{Optimizing for {Happiness} and {Productivity}:
  {Modeling} {Opportune} {Moments} for {Transitions} and {Breaks} at {Work}}.
  In \bibinfo{booktitle}{\emph{Proceedings of the 2020 {CHI} {Conference} on
  {Human} {Factors} in {Computing} {Systems}}} \emph{(\bibinfo{series}{{CHI}
  '20})}. \bibinfo{publisher}{Association for Computing Machinery},
  \bibinfo{address}{New York, NY, USA}, \bibinfo{pages}{1--15}.
\newblock
\showISBNx{978-1-4503-6708-0}
\urldef\tempurl%
\url{https://doi.org/10.1145/3313831.3376817}
\showDOI{\tempurl}


\bibitem[\protect\citeauthoryear{Kim, Jung, Ko, and Lee}{Kim
  et~al\mbox{.}}{2019b}]%
        {kim_goalkeeper_2019}
\bibfield{author}{\bibinfo{person}{Jaejeung Kim}, \bibinfo{person}{Hayoung
  Jung}, \bibinfo{person}{Minsam Ko}, {and} \bibinfo{person}{Uichin Lee}.}
  \bibinfo{year}{2019}\natexlab{b}.
\newblock \showarticletitle{{GoalKeeper}: {Exploring} {Interaction} {Lockout}
  {Mechanisms} for {Regulating} {Smartphone} {Use}}.
\newblock \bibinfo{journal}{\emph{Proceedings of the ACM on Interactive,
  Mobile, Wearable and Ubiquitous Technologies}} \bibinfo{volume}{3},
  \bibinfo{number}{1} (\bibinfo{date}{March} \bibinfo{year}{2019}),
  \bibinfo{pages}{16:1--16:29}.
\newblock
\urldef\tempurl%
\url{https://doi.org/10.1145/3314403}
\showDOI{\tempurl}


\bibitem[\protect\citeauthoryear{Kim, Park, Lee, Ko, and Lee}{Kim
  et~al\mbox{.}}{2019c}]%
        {kim_lockntype_2019}
\bibfield{author}{\bibinfo{person}{Jaejeung Kim}, \bibinfo{person}{Joonyoung
  Park}, \bibinfo{person}{Hyunsoo Lee}, \bibinfo{person}{Minsam Ko}, {and}
  \bibinfo{person}{Uichin Lee}.} \bibinfo{year}{2019}\natexlab{c}.
\newblock \showarticletitle{{LocknType}: {Lockout} {Task} {Intervention} for
  {Discouraging} {Smartphone} {App} {Use}}. In
  \bibinfo{booktitle}{\emph{Proceedings of the 2019 {CHI} {Conference} on
  {Human} {Factors} in {Computing} {Systems}}} \emph{(\bibinfo{series}{{CHI}
  '19})}. \bibinfo{publisher}{Association for Computing Machinery},
  \bibinfo{address}{New York, NY, USA}, \bibinfo{pages}{1--12}.
\newblock
\showISBNx{978-1-4503-5970-2}
\urldef\tempurl%
\url{https://doi.org/10.1145/3290605.3300927}
\showDOI{\tempurl}


\bibitem[\protect\citeauthoryear{Kim, Choe, Lee, and Seo}{Kim
  et~al\mbox{.}}{2019a}]%
        {kim_understanding_2019}
\bibfield{author}{\bibinfo{person}{Young-Ho Kim}, \bibinfo{person}{Eun~Kyoung
  Choe}, \bibinfo{person}{Bongshin Lee}, {and} \bibinfo{person}{Jinwook Seo}.}
  \bibinfo{year}{2019}\natexlab{a}.
\newblock \showarticletitle{Understanding {Personal} {Productivity}: {How}
  {Knowledge} {Workers} {Define}, {Evaluate}, and {Reflect} on {Their}
  {Productivity}}. In \bibinfo{booktitle}{\emph{Proceedings of the 2019 {CHI}
  {Conference} on {Human} {Factors} in {Computing} {Systems}}}
  \emph{(\bibinfo{series}{{CHI} '19})}. \bibinfo{publisher}{Association for
  Computing Machinery}, \bibinfo{address}{New York, NY, USA},
  \bibinfo{pages}{1--12}.
\newblock
\showISBNx{978-1-4503-5970-2}
\urldef\tempurl%
\url{https://doi.org/10.1145/3290605.3300845}
\showDOI{\tempurl}


\bibitem[\protect\citeauthoryear{Kuss and Pontes}{Kuss and Pontes}{2018}]%
        {kuss2018internet}
\bibfield{author}{\bibinfo{person}{Daria~J Kuss} {and}
  \bibinfo{person}{Halley~M Pontes}.} \bibinfo{year}{2018}\natexlab{}.
\newblock \bibinfo{booktitle}{\emph{Internet addiction}}.
  Vol.~\bibinfo{volume}{41}.
\newblock \bibinfo{publisher}{Hogrefe Verlag}.
\newblock


\bibitem[\protect\citeauthoryear{Lukoff, Yu, Kientz, and Hiniker}{Lukoff
  et~al\mbox{.}}{2018}]%
        {lukoff_what_2018}
\bibfield{author}{\bibinfo{person}{Kai Lukoff}, \bibinfo{person}{Cissy Yu},
  \bibinfo{person}{Julie Kientz}, {and} \bibinfo{person}{Alexis Hiniker}.}
  \bibinfo{year}{2018}\natexlab{}.
\newblock \showarticletitle{What {Makes} {Smartphone} {Use} {Meaningful} or
  {Meaningless}?}
\newblock \bibinfo{journal}{\emph{Proceedings of the ACM on Interactive,
  Mobile, Wearable and Ubiquitous Technologies}} \bibinfo{volume}{2},
  \bibinfo{number}{1} (\bibinfo{date}{March} \bibinfo{year}{2018}),
  \bibinfo{pages}{22:1--22:26}.
\newblock
\urldef\tempurl%
\url{https://doi.org/10.1145/3191754}
\showDOI{\tempurl}


\bibitem[\protect\citeauthoryear{Luo, Lee, Wohn, Rebar, Conroy, and Choe}{Luo
  et~al\mbox{.}}{2018}]%
        {luo_time_2018}
\bibfield{author}{\bibinfo{person}{Yuhan Luo}, \bibinfo{person}{Bongshin Lee},
  \bibinfo{person}{Donghee~Yvette Wohn}, \bibinfo{person}{Amanda~L. Rebar},
  \bibinfo{person}{David~E. Conroy}, {and} \bibinfo{person}{Eun~Kyoung Choe}.}
  \bibinfo{year}{2018}\natexlab{}.
\newblock \showarticletitle{Time for {Break}: {Understanding} {Information}
  {Workers}' {Sedentary} {Behavior} {Through} a {Break} {Prompting} {System}}.
\newblock In \bibinfo{booktitle}{\emph{Proceedings of the 2018 {CHI}
  {Conference} on {Human} {Factors} in {Computing} {Systems}}}.
  \bibinfo{publisher}{Association for Computing Machinery},
  \bibinfo{address}{New York, NY, USA}, \bibinfo{pages}{1--14}.
\newblock
\showISBNx{978-1-4503-5620-6}
\urldef\tempurl%
\url{https://doi.org/10.1145/3173574.3173701}
\showURL{%
\tempurl}


\bibitem[\protect\citeauthoryear{Lyngs, Lukoff, Slovak, Seymour, Webb, Jirotka,
  Zhao, Van~Kleek, and Shadbolt}{Lyngs et~al\mbox{.}}{2020}]%
        {lyngs2020just}
\bibfield{author}{\bibinfo{person}{Ulrik Lyngs}, \bibinfo{person}{Kai Lukoff},
  \bibinfo{person}{Petr Slovak}, \bibinfo{person}{William Seymour},
  \bibinfo{person}{Helena Webb}, \bibinfo{person}{Marina Jirotka},
  \bibinfo{person}{Jun Zhao}, \bibinfo{person}{Max Van~Kleek}, {and}
  \bibinfo{person}{Nigel Shadbolt}.} \bibinfo{year}{2020}\natexlab{}.
\newblock \showarticletitle{'I Just Want to Hack Myself to Not Get Distracted'
  Evaluating Design Interventions for Self-Control on Facebook}. In
  \bibinfo{booktitle}{\emph{Proceedings of the 2020 CHI Conference on Human
  Factors in Computing Systems}}. \bibinfo{pages}{1--15}.
\newblock


\bibitem[\protect\citeauthoryear{Mark, Czerwinski, and Iqbal}{Mark
  et~al\mbox{.}}{2018}]%
        {mark2018effects}
\bibfield{author}{\bibinfo{person}{Gloria Mark}, \bibinfo{person}{Mary
  Czerwinski}, {and} \bibinfo{person}{Shamsi~T Iqbal}.}
  \bibinfo{year}{2018}\natexlab{}.
\newblock \showarticletitle{Effects of individual differences in blocking
  workplace distractions}. In \bibinfo{booktitle}{\emph{Proceedings of the 2018
  CHI Conference on Human Factors in Computing Systems}}.
  \bibinfo{pages}{1--12}.
\newblock


\bibitem[\protect\citeauthoryear{Mark, Iqbal, and Czerwinski}{Mark
  et~al\mbox{.}}{2017}]%
        {mark2017blocking}
\bibfield{author}{\bibinfo{person}{Gloria Mark}, \bibinfo{person}{Shamsi
  Iqbal}, {and} \bibinfo{person}{Mary Czerwinski}.}
  \bibinfo{year}{2017}\natexlab{}.
\newblock \showarticletitle{How blocking distractions affects workplace focus
  and productivity}. In \bibinfo{booktitle}{\emph{Proceedings of the 2017 ACM
  International Joint Conference on Pervasive and Ubiquitous Computing and
  Proceedings of the 2017 ACM International Symposium on Wearable Computers}}.
  \bibinfo{pages}{928--934}.
\newblock


\bibitem[\protect\citeauthoryear{Meerkerk, Van Den~Eijnden, Vermulst, and
  Garretsen}{Meerkerk et~al\mbox{.}}{2009}]%
        {meerkerk2009compulsive}
\bibfield{author}{\bibinfo{person}{G-J Meerkerk}, \bibinfo{person}{Regina~JJM
  Van Den~Eijnden}, \bibinfo{person}{Ad~A Vermulst}, {and}
  \bibinfo{person}{Henk~FL Garretsen}.} \bibinfo{year}{2009}\natexlab{}.
\newblock \showarticletitle{The compulsive internet use scale (CIUS): some
  psychometric properties}.
\newblock \bibinfo{journal}{\emph{Cyberpsychology \& behavior}}
  \bibinfo{volume}{12}, \bibinfo{number}{1} (\bibinfo{year}{2009}),
  \bibinfo{pages}{1--6}.
\newblock


\bibitem[\protect\citeauthoryear{Meyer, Murphy, Zimmermann, and Fritz}{Meyer
  et~al\mbox{.}}{2017}]%
        {meyer_design_2017}
\bibfield{author}{\bibinfo{person}{Andre~N. Meyer}, \bibinfo{person}{Gail~C.
  Murphy}, \bibinfo{person}{Thomas Zimmermann}, {and} \bibinfo{person}{Thomas
  Fritz}.} \bibinfo{year}{2017}\natexlab{}.
\newblock \showarticletitle{Design {Recommendations} for {Self}-{Monitoring} in
  the {Workplace}: {Studies} in {Software} {Development}}.
\newblock \bibinfo{journal}{\emph{Proceedings of the ACM on Human-Computer
  Interaction}} \bibinfo{volume}{1}, \bibinfo{number}{CSCW}
  (\bibinfo{date}{Dec.} \bibinfo{year}{2017}), \bibinfo{pages}{79:1--79:24}.
\newblock
\urldef\tempurl%
\url{https://doi.org/10.1145/3134714}
\showDOI{\tempurl}


\bibitem[\protect\citeauthoryear{Okeke, Sobolev, Dell, and Estrin}{Okeke
  et~al\mbox{.}}{2018}]%
        {okeke_good_2018}
\bibfield{author}{\bibinfo{person}{Fabian Okeke}, \bibinfo{person}{Michael
  Sobolev}, \bibinfo{person}{Nicola Dell}, {and} \bibinfo{person}{Deborah
  Estrin}.} \bibinfo{year}{2018}\natexlab{}.
\newblock \showarticletitle{Good vibrations: can a digital nudge reduce digital
  overload?}. In \bibinfo{booktitle}{\emph{Proceedings of the 20th
  {International} {Conference} on {Human}-{Computer} {Interaction} with
  {Mobile} {Devices} and {Services}}} \emph{(\bibinfo{series}{{MobileHCI}
  '18})}. \bibinfo{publisher}{Association for Computing Machinery},
  \bibinfo{address}{New York, NY, USA}, \bibinfo{pages}{1--12}.
\newblock
\showISBNx{978-1-4503-5898-9}
\urldef\tempurl%
\url{https://doi.org/10.1145/3229434.3229463}
\showDOI{\tempurl}


\bibitem[\protect\citeauthoryear{Panero, Lane, and Napier}{Panero
  et~al\mbox{.}}{1997}]%
        {panero1997part}
\bibfield{author}{\bibinfo{person}{Jan~C Panero}, \bibinfo{person}{David~M
  Lane}, {and} \bibinfo{person}{H~Albert Napier}.}
  \bibinfo{year}{1997}\natexlab{}.
\newblock \showarticletitle{PART I: The Computer use Scale: Four Dimensions of
  how People use Computers}.
\newblock \bibinfo{journal}{\emph{Journal of Educational Computing Research}}
  \bibinfo{volume}{16}, \bibinfo{number}{4} (\bibinfo{year}{1997}),
  \bibinfo{pages}{297--315}.
\newblock


\bibitem[\protect\citeauthoryear{Popovich, Hyde, Zakrajsek, and
  Blumer}{Popovich et~al\mbox{.}}{1987}]%
        {popovich1987development}
\bibfield{author}{\bibinfo{person}{Paula~M Popovich}, \bibinfo{person}{Karen~R
  Hyde}, \bibinfo{person}{Todd Zakrajsek}, {and} \bibinfo{person}{Catherine
  Blumer}.} \bibinfo{year}{1987}\natexlab{}.
\newblock \showarticletitle{The development of the attitudes toward computer
  usage scale}.
\newblock \bibinfo{journal}{\emph{Educational and psychological measurement}}
  \bibinfo{volume}{47}, \bibinfo{number}{1} (\bibinfo{year}{1987}),
  \bibinfo{pages}{261--269}.
\newblock


\bibitem[\protect\citeauthoryear{Rooksby, Asadzadeh, Rost, Morrison, and
  Chalmers}{Rooksby et~al\mbox{.}}{2016}]%
        {rooksby_personal_2016}
\bibfield{author}{\bibinfo{person}{John Rooksby}, \bibinfo{person}{Parvin
  Asadzadeh}, \bibinfo{person}{Mattias Rost}, \bibinfo{person}{Alistair
  Morrison}, {and} \bibinfo{person}{Matthew Chalmers}.}
  \bibinfo{year}{2016}\natexlab{}.
\newblock \showarticletitle{Personal {Tracking} of {Screen} {Time} on {Digital}
  {Devices}}. In \bibinfo{booktitle}{\emph{Proceedings of the 2016 {CHI}
  {Conference} on {Human} {Factors} in {Computing} {Systems}}}
  \emph{(\bibinfo{series}{{CHI} '16})}. \bibinfo{publisher}{Association for
  Computing Machinery}, \bibinfo{address}{New York, NY, USA},
  \bibinfo{pages}{284--296}.
\newblock
\showISBNx{978-1-4503-3362-7}
\urldef\tempurl%
\url{https://doi.org/10.1145/2858036.2858055}
\showDOI{\tempurl}


\bibitem[\protect\citeauthoryear{Soliman~Elserty, Ahmed~Helmy, and
  Mohmed~Mounir}{Soliman~Elserty et~al\mbox{.}}{2020}]%
        {soliman2020smartphone}
\bibfield{author}{\bibinfo{person}{Noha Soliman~Elserty},
  \bibinfo{person}{Nesma Ahmed~Helmy}, {and} \bibinfo{person}{Khaled
  Mohmed~Mounir}.} \bibinfo{year}{2020}\natexlab{}.
\newblock \showarticletitle{Smartphone addiction and its relation to
  musculoskeletal pain in Egyptian physical therapy students}.
\newblock \bibinfo{journal}{\emph{European Journal of Physiotherapy}}
  \bibinfo{volume}{22}, \bibinfo{number}{2} (\bibinfo{year}{2020}),
  \bibinfo{pages}{70--78}.
\newblock


\bibitem[\protect\citeauthoryear{Subramani~Parasuraman, Yee, Chuon, and
  Ren}{Subramani~Parasuraman et~al\mbox{.}}{2017}]%
        {subramani2017smartphone}
\bibfield{author}{\bibinfo{person}{Aaseer Thamby~Sam Subramani~Parasuraman},
  \bibinfo{person}{Stephanie Wong~Kah Yee}, \bibinfo{person}{Bobby Lau~Chik
  Chuon}, {and} \bibinfo{person}{Lee~Yu Ren}.} \bibinfo{year}{2017}\natexlab{}.
\newblock \showarticletitle{Smartphone usage and increased risk of mobile phone
  addiction: A concurrent study}.
\newblock \bibinfo{journal}{\emph{International journal of pharmaceutical
  investigation}} \bibinfo{volume}{7}, \bibinfo{number}{3}
  (\bibinfo{year}{2017}), \bibinfo{pages}{125}.
\newblock


\bibitem[\protect\citeauthoryear{Wang and Reiterer}{Wang and Reiterer}{2019}]%
        {wang_point--choice_2019}
\bibfield{author}{\bibinfo{person}{Yunlong Wang} {and} \bibinfo{person}{Harald
  Reiterer}.} \bibinfo{year}{2019}\natexlab{}.
\newblock \showarticletitle{The {Point}-of-{Choice} {Prompt} or the
  {Always}-{On} {Progress} {Bar}? {A} {Pilot} {Study} of {Reminders} for
  {Prolonged} {Sedentary} {Behavior} {Change}}. In
  \bibinfo{booktitle}{\emph{Extended {Abstracts} of the 2019 {CHI} {Conference}
  on {Human} {Factors} in {Computing} {Systems}}} \emph{(\bibinfo{series}{{CHI}
  {EA} '19})}. \bibinfo{publisher}{Association for Computing Machinery},
  \bibinfo{address}{New York, NY, USA}, \bibinfo{pages}{1--6}.
\newblock
\showISBNx{978-1-4503-5971-9}
\urldef\tempurl%
\url{https://doi.org/10.1145/3290607.3313050}
\showDOI{\tempurl}


\end{thebibliography}

\end{document}